\documentclass[12pt,twoside]{article}
\usepackage[mathscr]{eucal}
\usepackage{amsmath,amsfonts,amssymb,amsthm,ulem,latexsym}

\usepackage{tabularx}
\usepackage{times}
\usepackage{cite}

\usepackage{multibox}

\usepackage{hyperref}

\def\be{\begin{equation}}
\def\ee{\end{equation}}
\def\ba{\begin{array}}
\def\ea{\end{array}}

\def\dps{\displaystyle}

\advance\textheight\topskip \textwidth 35.5pc \oddsidemargin 20pt
\renewcommand{\tilde}{\widetilde}
\renewcommand{\hat}{\widehat}
\newtheorem{prop}{Proposition}[section]

\newcommand{\assalgebra}{\mathscr}    

\renewcommand{\geq}{\,{\geqslant}\,}

\newcommand{\binner}[2]{%
  {\langle}\kern-4.15pt{\langle}#1{,}\,#2{\rangle}\kern-4.15pt{\rangle}}

\newcommand{\half}{\mathchoice{%
    \ffrac{1}{2}}{\frac{1}{2}}{\frac{1}{2}}{\frac{1}{2}}}

\newcommand{\ffrac}[2]{\raisebox{.5pt}%
  {\footnotesize$\displaystyle\frac{#1}{#2}$}\kern1pt}

\def\cN{\mathcal{N}}

\def\cZ{\mathcal{Z}}

\numberwithin{equation}{section} \makeatletter
\@addtoreset{equation}{section}

\makeatletter
\def\Appendix{\appendix
  \def\@seccntformat##1{Appendix~\csname the##1\endcsname.~~}}
\makeatother

\def\be{\begin{equation}}
\def\ee{\end{equation}}
\def\ba{\begin{array}}
\def\ea{\end{array}}

\def\dps{\displaystyle}

\def\ba{\begin{array}}
\def\ea{\end{array}}

\def\dps{\displaystyle}

\newcolumntype{x}[1]{%
>{\centering\hspace{0pt}}m{#1}}%
\newcolumntype{w}[1]{%
>{\raggedright\hspace{0pt}}m{#1}}%
\newcolumntype{z}[1]{%
>{\raggedleft\hspace{0pt}}m{#1}}%

\begin{document}

\begin{flushright}
FIAN-TD-2011-07 \\
\end{flushright}
\vspace{1cm}

\begin{center}

{\Large\textbf{Super Liouville conformal blocks
from $\bf{\cN=2}$ $\bf{SU(2)}$
\\ [8pt] 
quiver gauge theories}}

\vspace{.9cm}

{\large\textbf{V.~Belavin$^{1}$ and B.~Feigin$^{2,3}$}}

\vspace{0.5cm}

$^1$~\parbox[t]{0.85\textwidth}{\normalsize\it\raggedright
Theoretical Department, Lebedev Physical Institute, RAS, Moscow, Russia}
$^2$~\parbox[t]{0.85\textwidth}{\normalsize\it\raggedright
Landau Institute for Theoretical Physics, RAS, Chernogolovka, Russia}
$^3$~\parbox[t]{0.85\textwidth}{\normalsize\it\raggedright
Department of Mathematics, Higher School of Economics, Moscow, Russia}

\vspace{0.5cm}

\begin{abstract}
The conjecture about the correspondence between instanton partition functions in the $N=2$ SUSY Yang-Mills theory
and conformal blocks of two-dimensional conformal field theories is extended to the case of  
the $N=1$ supersymmetric conformal blocks. We find that the necessary modification of the moduli 
space of instantons requires additional restriction of $Z(2)$-symmetry. This leads to an explicit 
form of the $N=1$ superconformal blocks in terms of Young diagrams with two sorts of cells. 
\end{abstract}

\end{center}

{ \small \tableofcontents }

\section{Introduction}

In \cite{Alday:2009aq} Alday, Gaiotto and Tachikawa uncovered a relation between two-dimensional conformal 
field theories (CFT) and a certain class of  $\cN=2$ four-dimensional supersymmetric $SU(2)$ quiver gauge theories.  
In particular, it was argued that the conformal blocks~\cite{Belavin:1984vu} in the Liouville field theory 
coincide with the Nekrasov instanton partition functions. Further, this relation was generalized 
\cite{Alday:2010vg,Wyllard:2009hg,MM,Taki} to CFTs
with affine and  $\assalgebra{W}_k$-symmetry. It turned out that the extended $\assalgebra{W}_k$ conformal 
symmetry is related to the instanton counting for the $SU(k)$ gauge group. This development suggests that maybe 
for any type of chiral algebra there exist an explicit connection 
between conformal blocks (as well as other CFT ingredients) and some instanton partition functions.

The relation between affine algebras and the
geometry of instanton moduli varieties was realized quite long ago (see,
\textit{e.g.} \cite{Lusztig,Nakajima1,Nakajima}). In this context, $\assalgebra{W}$-algebras
arise from the so-called toroidal algebra, depending on several quantum
parameters, as a result of some special ``conformal'' limit. The
toroidal algebra acts on the cohomologies (equivariant K-theories) \cite{Feigin1} of
the instanton moduli spaces. In the conformal limit, this algebra reproduces
the coset of the form $\hat{gl}_k(n)/\hat{gl}_k(n-1)$, where $k$ denotes
the level of the current algebra. In the particular case where $k=2$,
this coset is isomorphic to $\assalgebra{H}\times\mathcal{W}_2$, where 
$\assalgebra{H}$ is the
Heisenberg algebra and $\assalgebra{W}_2$ is just the Virasoro algebra with
the central charge defined in terms of the parameter $n$. The AGT
relation corresponds to this situation. 

The varieties of the symmetric instanton moduli were studied
in~\cite{Nagao}. This is a subspace of the moduli space
consisting of fixed points under the action of some finite group. Once
the action of the finite group is introduced on the instanton moduli
space, the coset corresponding to the conformal limit changes. For
example, if the group is $\mathbb{Z}_2$, then the $\assalgebra{W}$-algebra
in the conformal limit is given by the coset
$\hat{gl}_k(n)/\hat{gl}_k(n-2)$. In particular, for $k=2$ this algebra
is isomorphic to $\hat{gl}_2(2)\times\assalgebra{NSR}$, where $\assalgebra{NSR}$ 
denotes the
Neveu--Schwarz--Ramond algebra. Because $\assalgebra{NSR}$ is the symmetry of the
$\cN=1$ super-Liouville field theory, it can be assumed that the
instanton calculus in this particular case can be related to the $\cN=1$
super Liouville conformal blocks. Apparently, the higher cyclic groups
$\mathbb{Z}_m$ may correspond to the parafermionic conformal field
theories. We are focusing on the $\mathbb{Z}_2$ case here.

Our main result can be summarized as follows. We consider the
two-dimensional $\cN=1$ superconformal field theory. We show that the
conformal block in the Whittaker limit is related to the instanton
partition function of the $SU(2)$ Yang--Mills theory evaluated on the
$\mathbb{Z}_2$ symmetric instanton moduli space. The $\mathbb{Z}_2$ symmetry
reduces the moduli space and modifies the instanton partition function.
This relation gives a new explicit representation of the $\cN=1$ 
superconformal block function in terms of Young diagrams with two sorts of
cells. In this paper, we only treat pure gauge theories. Theories with 
matter fields will be studied elsewhere.  

The paper is organized as follows. In section \ref{Bosonic} we briefly recall the 
results of the AGT conjecture in the ordinary Liouville case.
Section \ref{N1CFT} is devoted to description of the $\cN=1$ SUSY conformal field theory and the construction
of conformal blocks via the standard bootstrap approach~\cite{Belavin:1984vu}. In section \ref{ADHM}
we briefly review the localization method \cite{Flume:2002az,B,BGV} based on the equivariant form of the moduli 
integral. This consideration  
leads to Nekrasov's results for the instanton partition function~\cite{Nekrasov:2002qd}. 
Section \ref{ADHMZ2} is the main part of the present paper. There we consider 
the structure of the modified moduli space corresponding to the $\cN=1$ super Liouville theory.
Then we derive an expression of $N$-instantons
contribution to the partition function in terms of the colored Young diagrams and formulate our conjecture
about relations between the modified instanton partition function and the $\cN=1$ super Liouville conformal block. 
We verify the analytic properties of 
the new representation for the conformal block and perform some lowest levels checks in Section \ref{checks}. 
In the Conclusion we give a brief summary and discuss some open problems.

\section{AGT conjecture}\label{Bosonic}

To illustrate AGT correspondence we consider the four-point conformal blocks on a sphere 
associated with four primary fields $\Phi_{\Delta_i}$ of conformal weights $\Delta_i$. 
This section mainly serves to set our conventions and notation.

The AGT conjecture states the equality between Nekrasov's partition function and the Liouville conformal block. 
For the $4$-point functions we have
the following
\begin{equation}
\label{AGT-conjecture}
\sum_{N=0}^\infty q^N \sum_{Y_1, Y_2}^{|Y_1| +|Y_2|=N} \cZ_{Y_1,Y_2}
\;\;\; \stackrel{AGT}{=}
\;\;\;(1-q)^{{2(\frac{Q}{2}+\lambda_1)(\frac{Q}{2}-\lambda_3)}} F(\Delta_1, \Delta_2, \Delta_3, \Delta_4|\Delta|q),
\end{equation}
where the function $F = F(\Delta_i|\Delta|q)$, $i=1,2,3,4$ on the right hand side
is the conformal block of the Liouville theory with the central charge $c=1+6Q^2$, while the so called background
charge $Q$ is related to the Liouville coupling constant $b$ as $Q=b+b^{-1}$. In addition to the central charge, the
four-point conformal block depends on the four-point projective invariant $q$, four ``external'' dimensions $\Delta_i$
and the ``intermediate''
dimension $\Delta$ 
\begin{equation}
\label{dims}
\Delta=\frac{Q^2}{4}-P^2,\qquad \Delta_i=\frac{Q^2}{4}-\lambda_i^2.
\end{equation}
In the framework of the bootstrap approach~\cite{Belavin:1984vu} the conformal block function is defined as follows
\begin{equation}
F(\Delta_i|\Delta|q)= \sum_{N=0}^\infty q^N \;  {}_{12}\langle N| N\rangle_{34}  \;,
\end{equation}
where the so-called ``chain vectors'' $| N\rangle_{12} \equiv | N\rangle_{\Delta_1 \Delta_2}$ are given
in terms of the Virasoro generators
$L_k$ and are built by using the following recursive relations
\be
L_k | N\rangle_{\Delta_1 \Delta_2} = (\Delta +k\Delta_1 - \Delta_2 +N-k) | N-k\rangle_{\Delta_1 \Delta_2}\;,
\label{boson-recur}
\ee
for any $k>0$.

The function $\cZ_{Y_1,Y_2}$ in~\eqref{AGT-conjecture}
is given by
\be
\cZ_{Y_1,Y_2} = \frac{Z_\text{\sf{f}}(\vec{a},\vec{Y},\mu_1) Z_\text{\sf{f}}(\vec{a},\vec{Y},\mu_2)
 Z_{\text{\sf{af}}}(\vec{a},\vec{Y},\mu_3) Z_{\text{\sf{af}}}(\vec{a},\vec{Y},\mu_4)}
{Z_{\text{\sf{vec}}}(\vec{a},\vec{Y})}\;.
\ee
By $\vec{Y}$, $\vec{a}$, ...  we denote pairs $(Y_1, Y_2)$, $(a_1, a_2)$, etc. The explicit form of the functions $Z_{\text{\sf{f}}}(\vec{a},\vec{Y},\mu)$,
$Z_{\text{\sf{af}}}(\vec{a},\vec{Y},\mu)$ and $Z_{\text{\sf{vec}}}(\vec{a},\vec{Y},\mu)$ are
\begin{equation}
\label{Zf}
\begin{aligned}
Z_{\text{\sf{f}}}(\vec{a},\vec{Y},\mu)=\prod_{i=1}^2\prod_{s \in Y_i}(\phi(a_i,s)-\mu+Q),
\end{aligned}
\end{equation}
\begin{equation}
\label{Zaf}
\begin{aligned}
Z_{\text{\sf{af}}}(\vec{a},\vec{Y},\mu)=\prod_{i=1}^2\prod_{s \in Y_i}(\phi(a_i,s)+\mu),
\end{aligned}
\end{equation}
where $\phi(a,s)$ is given by \eqref{YTphi} and
\begin{equation}
\label{ZfZvec}
\begin{aligned}
Z_{\text{\sf{vec}}}(\vec{a},\vec{Y})=\prod_{i,j=1}^{2}
    \prod_{s\in Y_{i}}E\bigl(a_{i}-a_{j},Y_i,Y_j\bigl|s\bigr)
\,\,\prod_{i,j=1}^{2}
    \prod_{s\in Y_{i}}(Q-E\bigl(a_{i}-a_{j},Y_i,Y_j\bigl|s\bigr)),
\end{aligned}
\end{equation}
where $E\bigl(a_i-a_j,Y_i,Y_j\bigl|s\bigr)$ is defined by \eqref{YTE}. It is assumed that
 factors in the above products associated with $Y = \varnothing$ are set to $1$.

\paragraph{Integer partitions.}  We operate with integer partitions $Y = [k_1, ...,
k_m]$ where integers are ordered as in \eqref{ordering}. A
particular partition can be visualized as a Young diagram disposed
in one or another way. A somewhat standard choice is to adjust
$k_i$ to horizontal rows. It is called symmetric basis and the
respective diagram is the following plaquette\footnote{To
reproduce the AGT convention one should rotate this diagram
counterclockwise to the angle $\pi/2$ and rename $k_i =
\lambda_i$.}

\be
\label{YT1}
\begin{picture}(75,70)(0,2)
\multiframe(0,0)(10.5,0){1}(10,10){}\put(20,0){$k_m$}\put(0,70){$k^\prime_1$}
\multiframe(0,10.5)(10.5,0){2}(10,10){}{}\put(25,12.5){$k_{m-1}$}
\put(-0.2,20){\line(0,1){25}}
\put(9,26){$\vdots$}
\multiframe(0,41.5)(10.5,0){5}(10,10){}{}{}{}{} \put(60,42.5){$k_2$}
\multiframe(0,52)(10.5,0){7}(10,10){}{}{}{}{}{}{} \put(79,55){$k_1$}
\end{picture}
\ee

\vspace{.2cm}
Let us introduce transposed Young diagram $Y^{T} = [k'_1, ...., k'_{l}]$ associated with $Y$.
It is given just by passing to antisymmetric basis, \textit{i.e.},
ordered integers $k'_i$ are adjusted to columns. The leftmost column is of height
$k'_1$.

A cell $s\in Y$ has coordinates $(i,j)$ such that $i$ and $j$ label a respective row and a column.
Functions $\phi(a,s)$ and $E\bigl(a,Y_1,Y_2\bigl|s\bigr)$ are
defined as follows
\be
\label{YTphi}
\phi(a,s)=a+b(i-1)+b^{-1}(j-1)\;,
\ee
\be
\label{YTE}
E\bigl(a,Y_1,Y_2\bigl|s\bigr)=
a+b(L_{\scriptscriptstyle{Y_1}}(s)+1)-b^{-1} A_{\scriptscriptstyle{Y_2}}(s)\;,
\ee
where arm-length function $A_{\scriptscriptstyle{Y}}(s)$ and leg-length function  $L_{\scriptscriptstyle{Y}}(s)$
for a cell $s\in Y$ are given by
\be
\label{legsarms}
A_{\scriptscriptstyle{Y}}(s) = k_i-j\;,
\qquad
L_{\scriptscriptstyle{Y}}(s) = k'_j - i\;.
\ee

The Nekrasov's partition function parameters are related to the parameters of the conformal block
$\Delta_i$, $\Delta$ and $c$ as follows:
\begin{equation}
\begin{aligned}[centered]
\mu_1=\frac Q2-(\lambda_1+\lambda_2),\qquad \mu_2=\frac Q2-(\lambda_1-\lambda_2), \\
\mu_3=\frac Q2-(\lambda_3+\lambda_4),\qquad \mu_4=\frac Q2-(\lambda_3-\lambda_4), \\
\end{aligned}
\label{parameters-mu}
\end{equation}
and
\begin{equation}
\vec{a}=(a,-a),\qquad a=P.
\label{parameter-a}
\end{equation}

\paragraph{Whittaker vector.}

In \cite{WhittakGaiot,MMM} several degenerated versions of the AGT conjecture were proposed.
In particular it was shown that the norm of the Whittaker vector \cite{Kostant} coincides with 
the Nekrasov partition function for pure gauge theory. In what follows we are dealing with this particular
case. Whittaker vector is defined as follows
\begin{equation}
V=\sum_{N=0}^{\infty} q^N |N\rangle,
\end{equation}
where $|N\rangle$ satisfies
\begin{equation}
\begin{cases}
L_0|N\rangle=(\Delta+N)|N\rangle,\\
L_1 |N\rangle=|N-1\rangle,\\
L_k |N\rangle=0 \,\,\,\, \text{for}\,\,\,\,k>1.
\end{cases}
\label{BosonWhitt}
\end{equation}
Let us find the coefficients of the Whittaker vector 
\begin{equation}
|N\rangle=\sum_{Y,|Y|=N} \beta_{Y} |Y\rangle,
\end{equation}
where $Y$ denotes the standard  basis in the Verma module.
This decomposition implies
\begin{eqnarray}
\beta_Y=(M^{-1})_{YY'} \langle Y'|N\rangle,
\label{beta}
\end{eqnarray}
where $M^{-1}$ is inverse of the scalar product matrix $M_{Y',Y}=\langle Y'|Y\rangle$.
From~\eqref{BosonWhitt} it follows that
\begin{equation}
\langle Y'|N\rangle=\delta_{Y',1^N}\;,
\end{equation}
and
\begin{eqnarray}
\beta_{Y}=(M^{-1})_{Y,1^N}\;.
\end{eqnarray}
Thus, the norm of the Whittaker vector is given by the scalar product
\begin{equation}
\langle N|N\rangle=\sum_{Y,|Y|=N}\beta_Y\langle N|Y\rangle=
\sum_{Y,|Y|=N}\beta_Y \beta_{1^N} \delta_{Y,1^N}=[(M^{-1})_{1^N,1^N} ]^2.
\end{equation}
One can easily see that in the limit $\Delta_{1,2}\rightarrow\infty$, after appropriate
rescaling of the chain vectors, the recursive relations \eqref{boson-recur} reproduces \eqref{BosonWhitt}.  
So that the norm of the Whittaker vector for $\Delta=\Delta(a)$ is related to the corresponding limit of the 
four-point conformal block
\begin{equation}
\langle N|N\rangle=\sum_{\vec{Y},|Y|=N} \frac{1}{Z_{\text{\sf{vec}}}(\vec{a},\vec{Y})}.
\end{equation}

\section{Super Liouville field theory}\label{N1CFT}
In this section we recall some details about Super Liouville field theory (SLFT) \cite{Polyakov,Belavin:2007gz} necessary for the 
forthcoming discussion. The Lagrangian of the theory reads
\begin{equation}
\mathcal{L}_{\text{SLFT}}=\frac1{8\pi}\left(  \partial_{a}\phi\right)
^{2}+\frac1{2\pi}\left(  \psi\bar\partial\psi+\bar\psi\partial\bar\psi\right)
+2i\mu b^{2}\bar\psi\psi e^{b\phi}+2\pi b^{2}\mu^{2}e^{2b\phi},
\label{SL}
\end{equation}
where the
scale parameter $\mu$ is called the cosmological constant and the coupling constant $b$ is related 
through the ``background charge'' $Q=b^{-1}+b$ to the central charge
\begin{equation}
c=1+2Q^{2}\label{cQ}%
\end{equation}
of the Neveu-Schwarz-Ramond ($\assalgebra{NSR}$) algebra
\be
\label{NRS}
\ba{l}
\dps
[L_m, L_n] = (n-m) L_{n+m} + \frac{c}{8}(n^3-n)\delta_{n+m}\;,
\\
\\
\dps
\{G_r, G_s\} = 2L_{r+s}+\frac{1}{2} c(r^2-\frac{1}{4})\delta_{n+m}\;,
\\
\\
\dps
[L_n, G_r] = (\frac{1}{2}n-r)G_{n+r}\;.
\ea
\ee
We will consider the NS sector defined by
\be
r,s \in \mathbb{Z} +\frac{1}{2}\;.
\ee
The states of the respective superconformal module are build as an ordered
\be
\label{susyvec}
|{\Delta,Y}\rangle = L_{-k_1} ... L_{-k_m} G_{-r_1} ... G_{-r_n} |\Delta\rangle\;,
\ee
where the highest weight vector $|\Delta\rangle$  is annihilated by all positive-frequency
generators and has the conformal dimension $\Delta$ defined by $L_0 |\Delta\rangle = \Delta |\Delta\rangle$.
Vectors $|\Delta\rangle$ and $G_{-1/2} |\Delta\rangle$ form primary superdublet which is denoted 
$(\Phi_{\Delta},\Psi_{\Delta})$ and $L_0 \Psi_{\Delta}=(\Delta+1/2)\Psi_{\Delta}$. We parametrize 
the conformal dimension as follows
\begin{equation}
\Delta(\lambda)=\frac{Q^2}{8}-\frac{\lambda^2}{2}. 
\end{equation}
Label $Y$ denotes a partition of some (half-)integer number $N$ 
\begin{equation}
Y = [k_1, ..., k_m | r_1, ... , r_n]
\end{equation}
such that
\be
\label{ordering}
k_1 \geq k_2 \geq  ... \geq k_m\;,
\quad
r_1 > r_2 >  ... > r_n\;,
\qquad
\sum_{i=1}^m k_i + \sum_{j=1}^n r_j = N\;.
\ee
The value of $N = \dps 0,\half, 1, ...$ fixes a particular level in the superconformal module.

\paragraph{$\cN=1$ conformal blocks.} The conformal block functions of the $\cN=1$ super Liouville theory
were intensively studied in the series of papers 
\cite{Hadasz:2006sb,Belavin:2007zz,Belavin:2007gz,Belavin:2007eq,Hadasz:2007nt}.
The  $4$-point correlation function of bosonic primaries $\Phi_i$ with
conformal weights $\Delta_i$ is given by
\be
\ba{c}
\dps
\langle \Phi_1(q) \Phi_2(0) \Phi_3(1) \Phi_4(\infty) \rangle  =(q\bar q)^{\Delta - \Delta_1 - \Delta_2} \sum_\Delta \Big( C^\Delta_{12}C^\Delta_{34} F_0(\Delta_i|\Delta|q)F_0(\Delta_i|\Delta|\bar q)
\\
\\
\dps
\hspace{6cm} + \tilde C^\Delta_{12}\tilde C^\Delta_{34} F_1(\Delta_i|\Delta|q)F_1(\Delta_i|\Delta|\bar q)\Big)\;.
\ea
\ee
The superconformal blocks $F_{0,1}$ have form
\begin{equation}
\begin{aligned}
\label{superblock}
&F_0(\Delta_i|\Delta|q) = \sum_{N=0,1,...} q^N F^{(N)}(\Delta_i|\Delta)\;,\\
&F_1(\Delta_i|\Delta|q) = \sum_{N=1/2,3/2,...} q^N F^{(N)}(\Delta_i|\Delta)\;,
\end{aligned}
\end{equation}
where
\be
\label{superblockcoef}
F^{(N)}(\Delta_i|\Delta) = {}_{12}\langle N| N\rangle_{34}
\ee
and vectors $| N\rangle_{12}$ for $N=0,1/2, 1,  ...$ are defined in terms of $\assalgebra{NSR}$ generators 
as linear combinations on the Nth level arising in the operator product expansion 
$\Phi_1(q)\Phi_2(0)$. 
They satisfy the following recursive relations 
\begin{equation}
\begin{cases}
G_k|N\rangle_{12}={\widetilde{|N-k \rangle}}_{12},\\
G_k{\widetilde{|N\rangle}}_{12}=
[\Delta+2k\Delta_1-\Delta_2+N-k]|N-k\rangle_{12},\end{cases}
\label{chain}
\end{equation}
where parameter $k$ runs over  half-integer values, $k=\frac{1}{2}, \frac{3}{2},...\,$
and ${\widetilde{|N\rangle}}_{12}$ is the contribution of Nth level descendents in the operator product expansion
$\Psi_1(q)\Phi_2(0)$.
\paragraph{Supersymmetric Whittaker vector.}
After appropriate re-scaling the chain vectors,
the limiting procedure $\Delta_{1,3}\rightarrow \infty$ for  \eqref{chain}
yields the following recursive  equations
\be
\ba{l}
\dps
G_{\half } |N\rangle ={\widetilde{|N-\half\rangle}}\;,
\qquad
G_r|N\rangle = 0\;,
\quad  r>\half\;,
\\
\\
\dps
G_{\half} {\widetilde{|N\rangle}} =|N-\half\rangle\;,
\qquad
G_r{\widetilde{|N\rangle}}=0\;,
\quad r>\half\;.

\ea
\ee
In what follows we are interested in the study of the conformal block function in the Whittaker limit
\begin{equation}
\begin{aligned}
\label{whittsuperblock}
&F_0(\Delta|q) = \sum_{N=0,1,...}\,\,\, q^N \langle N| N\rangle\;,\\
&F_1(\Delta|q) = \sum_{N=1/2,3/2,...} q^N \langle N| N\rangle\;.
\end{aligned}
\end{equation}
Here we list the few lowest coefficients
\begin{equation}
\begin{aligned}
&\langle 0\,|\,0\rangle=1,\\
&\langle \frac 12|\frac 12\rangle=\frac{1}{2\Delta},\\
&\langle 1\,|\, 1\rangle=\frac{1}{8\Delta},\\
&\langle \frac 32|\frac 32\rangle=\frac{c+2\Delta}{8\Delta(c-6\Delta+2 c \Delta+4\Delta^2)},\\
&\langle 2\,|\,2\rangle=\frac{3c+3c^2-34\Delta+22 c \Delta+32\Delta^2}
{64\Delta(-3+3c+16\Delta)(c-6\Delta+2 c\Delta+4\Delta^2)},\\
&\langle \frac 52\,|\,\frac 52\rangle=
\frac{-27 c+42c^2+9c^3+2\Delta+50 c \Delta +72 c^2 \Delta-228 \Delta^2+140 c \Delta^2+64 \Delta^3}
{128 \Delta(-3+3c+16\Delta)(5+3 c-11\Delta+3 c\Delta+2\Delta^2)(c-6\Delta+2c \Delta+4 \Delta^2)}.
\label{lowestcoef}
\end{aligned}
\end{equation}

\section{ADHM construction and the determinants of the vector field}\label{ADHM}

In~\cite{Nekrasov:2002qd, Flume:2002az} the form of $\cN=2$ $SU(k)$  
instanton partition function (in what follows we are dealing with $SU(2)$ case) was derived 
as an integral of  the equivariantly form, which is defined in terms of the 
vector field $v$ acting on the moduli space $\mathcal{M}_N$ ($N$ is the topological charge). 
This action will be specified below. 
By means of the localization technique \cite{B}, \cite{BGV}, the evaluation of the 
moduli integral is reduced to the calculation of the determinants \cite{Nekrasov:2002qd,Flume:2002az,Nakajima} of the vector field $v$ in the vicinity of fixed points 
\begin{eqnarray}
\mathcal{Z}_N \left(a, \epsilon_1, \epsilon_2 \right)=  \sum_{n} \frac{1}{\det_n v}. \label{localization}
\end{eqnarray}
Here $n$ numerates fixed points of the vector field. 
We quote the ADHM data \cite{adhm}  for the construction of $SU(2)$ instantons 
(see also \cite{BPSz,BZakh,aw,dm,DKM'}). These data consist of complex
matrices, two $N\times N$ matrices $B_1$, $B_2$, a $N\times 2$
matrix $I$ and a $2\times N$ matrix $J$, fulfilling a certain regularity
condition \cite{adhm} and obeying the relations
\begin{eqnarray}
\label{complexADHM}
\left[ B_1, B_2\right]+IJ=0, \\
\label{realADHM} \left[ B_1, B_1^{\dagger}\right]+\left[ B_2, B_2^{\dagger}\right]+II^{\dagger}-J^{\dagger}J=0,
\end{eqnarray}
where $"^{\dagger}"$ denotes hermitian conjugation. The regularity condition claims that 
$N$-di\-men\-si\-onal complex space $\mathbb{C}^N$ is spanned entirely by  
the repeated action of $B_1$ and $B_2$ on the column vectors $I_{1,2}$. 
The data are redundant in the sense that sets of matrices
related by $U(N)$ transformations,
\begin{eqnarray}
B_i^{\prime}=gB_ig^{-1}, \, \, \, I^{\prime}=gI, \, \, \, J^{\prime}=Jg^{-1}; \, \, \, \, g\in U(N)
\label{gauge}
\end{eqnarray}
are equivalent and represent the same point 
in  $\mathcal{M}_N$ (i.e. give rise to the same Yang-Mills configuration).  

The action of the vector field $v$ on the ADHM data is given by
\begin{equation}
B_l\rightarrow t_l B_l ; \, \, \, \, I\rightarrow It_v; \, \, \,
\, J\rightarrow t_1 t_2 t_v^{-1}J,
\label{combinedaction}
\end{equation}
where parameters $t_l\equiv \exp \epsilon_l \tau$, $l=1,2$ and $t_v=\exp a \sigma_3 \tau$.

Fixed points are defined by the
conditions:
\begin{equation}
t_l B_l=g^{-1}B_lg ; \, \, \, \, It_v=g^{-1}I; \, \, \, \, t_1 t_2
t_v^{-1}J=J g. \label{fixedpointconditions}
\end{equation}
The solutions of this system can be parameterized by pairs of Young diagrams $(Y_1,Y_2)$ such that the total number 
of cells $|Y_1|+|Y_2|=N$. This comes from the observation that there should exist $N$ linear independent vectors 
of the form 
$B_1^{i_1} B_2^{j_1} I_1$ and  $B_1^{i_2} B_2^{j_2} I_2$ which are the eigenvectors of the matrix $g$.
These vectors correspond to the cells $(i_1,j_1)\in Y_1$ and $(i_2,j_2)\in Y_2$ respectively.
The structure of the Young diagram just reflects the special way of ordering of the vectors.
It is convenient to use them as a basis in $\mathbb{C}^N$, then the explicit form of the ADHM date is defined 
straightforwardly  
\begin{equation}
\begin{aligned}
&g_{ss'}=\delta_{ss'} t_1^{i_s-1}t_2^{j_s-1},\\
&(B_1)_{ss'}=\delta_{i_s+1,i_{s'}} \delta_{j_s,j_{s'}},\\
&(B_2)_{ss'}=\delta_{i_s,i_{s'}} \delta_{j_{s+1},j_{s'}},\\
&(I_1)_{s}=\delta_{s,1},\\
&(I_2)_{s}=\delta_{s,|Y_1|+1},\\
&J=0,
\end{aligned}
\end{equation} 
where $s=(i_s,j_s)$.

To evaluate the determinant of the vector field one needs to find all eigenvectors
of the vector field on the tangent space passing through the fixed points
\begin{equation}
\begin{aligned}
&t_i \delta B_i=\Lambda\, g \delta B_i g^{-1},\\
&\delta I t =\Lambda\, g \delta I,\\
&t_1 t_2 t^{-1} \delta J=\Lambda\, \delta J g^{-1}.
\end{aligned}
\end{equation}
This is equivalent to the following set of equations
\begin{equation}
\begin{aligned}
&\lambda\, (\delta B_i)_{s s'}=(\epsilon_i+\phi_{s'}-\phi_s)\, (\delta B_i)_{ss'},\\
&\lambda\, (\delta I)_{s p} =(a_p-\phi_s)\,  (\delta I)_{s p},\\
&\lambda\, (\delta J)_{p s}=(\epsilon_1+\epsilon_2-a_p+\phi_s)\, (\delta J)_{p s},
\label{linsyst}
\end{aligned}
\end{equation}
where $\Lambda=\exp \lambda\tau$, $g_{ss}=\exp \phi_s \tau$ and 
\begin{equation}
\phi_s=(i_s-1)\epsilon_1+(j_s-1)\epsilon_2+a_{p(s)}.
\end{equation}
System \eqref{linsyst} gives all possible eigenvectors of the vector field. We should keep only those 
which belong to the tangent space. Essentially this means excluding variations breaking
ADHM constraints. On the Moduli space
\begin{eqnarray}
\label{complexvar}
&\left[\delta B_1, B_2\right]+[B_1,\delta B_2]+\delta I J+ I \delta J=0,\\
\label{realvar}&\left[\delta B_l,B_l^{\dagger}\right]+[B_l,\delta B_l^{\dagger}]+\delta I I^{\dagger}+I \delta I^{\dagger}-
\delta J^{\dagger} J- J^{\dagger} \delta J=0.
\end{eqnarray}
Gauge symmetry can be taken into account in the following manner.
We fix a gauge in which $\delta B_{1,2}, \delta I,\delta J$ are orthogonal to any gauge transformation of $B_{1,2},I,J$.
This gives additional constraint 
\begin{equation}
\label{gaugevar}
\left[\delta B_l,B_l^{\dagger}\right]-[B_l,\delta B_l^{\dagger}]+\delta I I^{\dagger}-I \delta I^{\dagger}+
\delta J^{\dagger} J- J^{\dagger} \delta J=0.
\end{equation}
We note that \eqref{realvar} and \eqref{gaugevar} are the real and the imaginary parts of the following 
equation
\begin{equation}
\label{realgaugevar}
\left[\delta B_l,B_l^{\dagger}\right]+\delta I I^{\dagger}- J^{\dagger} \delta J=0.
\end{equation}
The variations in the LHS of \eqref{complexvar} and \eqref{realgaugevar} should be excluded 
from \eqref{linsyst}. The corresponding eigenvalues are defined from the equations 
\begin{equation}
\begin{aligned}
&t_1 t_2 (\left[\delta B_1, B_2\right]+[B_1,\delta B_2]+\delta I J+ I \delta J)=
\Lambda\, g \bigg(\left[\delta B_1,B_2\right]+[B_1,\delta B_2]+\delta I J+ I \delta J\bigg) g^{-1},\\
&\left[\delta B_l,B_l^{\dagger}\right]+\delta I I^{\dagger}- J^{\dagger} \delta J=\Lambda\,g\bigg(\left[\delta B_l,B_l^{\dagger}\right]+\delta I I^{\dagger}- J^{\dagger} \delta J\bigg)g^{-1}.
\end{aligned}
\end{equation}
One finds the following eigenvalues, which should be excluded from~\eqref{linsyst}: 
\begin{equation}
\begin{aligned}
&\lambda=(\epsilon_1+\epsilon_2+\phi_s-\phi_{s'}),\\
&\lambda=(\phi_s-\phi_{s'}).
\end{aligned}
\end{equation}
Thus, the determinant of the vector field~\eqref{combinedaction} is given by
\begin{equation}
\det v=\frac{\prod_{s,s'\in \vec{Y}}(\epsilon_1+\phi_{s'}-\phi_s)(\epsilon_2+\phi_{s'}-\phi_s)
\prod_{l=1,2; s  \in \vec{Y}}(a_l-\phi_{s})(\epsilon_1+\epsilon_2- a_l+\phi_{s})}
{\prod_{s,s' \in \vec{Y}}(\phi_{s'}-\phi_s)(\epsilon_1+\epsilon_2-\phi_{s'}+\phi_s)}
\end{equation}
Re-expressed in terms of arm-length and leg-length this expression gives \eqref{ZfZvec}
\begin{equation}
Z_{\text{vec}}=\det v,
\end{equation}
once $\epsilon_1=b^{-1}$ and $\epsilon_2=b$.

\section{Modified Moduli space and super conformal blocks}\label{ADHMZ2}

We define the subspace of the Moduli space $\mathcal{M}_{\text{sym}}$ for $SU(2)$ gauge group obtained by the 
following additional restriction of $\mathbb{Z}_2$ symmetry
\begin{equation}
-B_{1,2}=P B_{1,2} P^{-1}; \qquad I = P I;\qquad J=J P^{-1}.
\label{Zsym}
\end{equation}
Here $P\in U(N)$ is some gauge transformation. We suggest to consider the following proposition.
\begin{prop}
The $2N$-instanton contribution to the moduli integral, evaluated on the
$\mathbb{Z}_2$ symmetric subspace of the moduli space
$\mathcal{M}_{\text{sym}}$, reproduces the $N$th-level conformal block
coefficients in the Neveu--Schwarz sector of the $\cN=1$ super Liouville
theory up to some factor related to
$\hat{gl}_2(2)$.
\end{prop}

First we note that  $\mathcal{M}_{\text{sym}}$ contains all fixed points of the vector
field \eqref{combinedaction} found in the previous section. Indeed, from \eqref{Zsym}
one finds the explicit action of $P$ on the basis vectors 
\begin{equation}
P(B_1^{i-1}B_2^{j-1} I_{\alpha})=(-1)^{i+j} B_1^{i-1}B_2^{j-1} I_{\alpha},
\end{equation}
so that the matrix elements are given explicitly, $P_{ss'}=(-1)^{i_s+j_s}\delta_{ss'}$. Below we denote 
$P(s)=(-1)^{i_s+j_s}$. A new feature in comparison with the results in the preceding section is
this $P$-characteristic assigned to each cell in the Young diagrams
related to the fixed points. To visualize this property, we use the
convention that a cell with coordinates of the same or different
parities are respectively white or black, as if we wrote the Young
diagrams on a chess board. Then $P(s)=1$ for white cells and $P(s)=-1$ for black ones.
Consequently, the fixed points can be classified by the number of white and black cells, $N_{+}$ and $N_{-}$.
This reflects the new structure of the manifold $\mathcal{M}_{\text{sym}}$ as a disjoint union of 
components $\mathcal{M}_{\text{sym}}(N_{+},N_{-})$. Each component is connected and can be considered separately.

Now we consider the action of the vector field~\eqref{combinedaction} in $\mathcal{M}_{\text{sym}}$. The tangent 
space is reduced by the additional requirement~\eqref{Zsym}
\begin{equation}
-\delta B_{1,2}=P \delta B_{1,2} P^{-1}; \qquad \delta I = P \delta I;\qquad \delta J=\delta J P^{-1},
\label{Zsymvar}
\end{equation}
or, on the level of the matrix elements,
\begin{equation}
-(\delta B_{1,2})_{ss'}=P(s) (\delta B_{1,2})_{ss'} P(s'); 
\quad (\delta I)_{sp} = P(s) (\delta I)_{sp};\quad (\delta J)_{ps}=(\delta J)_{ps} P(s),
\label{Zsymvarmatrix}
\end{equation}
The first relation in \eqref{Zsymvarmatrix} means that only eigenvectors $(\delta B_{1,2})_{ss'}$
with the different colors of $s$ and $s'$  belong to $Z_{\text{sym}}$. Similarly,
the second one leaves $(\delta J)_{ps}$ only if $s$ is white. The variations, which should be excluded
\eqref{complexvar} and \eqref{realgaugevar} belong to $\mathcal{M}_{\text{sym}}$ only for the matrix elements
between the states of the same color. Thus, we get the new determinant of the
vector field \eqref{combinedaction}
\begin{equation}
\begin{aligned}
\det{}' v=
\frac{\prod_{\substack{s,s'\in \vec{Y}\\P(s)\neq P(s')}}(\epsilon_1+\phi_{s'}-\phi_s)(\epsilon_2+\phi_{s'}-\phi_s)
\prod_{\substack{\alpha=1,2; s  \in \vec{Y}\\P(s)=1}}(a_{\alpha}-\phi_{s})
(\epsilon_1+\epsilon_2- a_{\alpha}+\phi_{s})}
{\prod_{\substack{s,s' \in \vec{Y}\\P(s)=P(s')}}(\phi_{s'}-\phi_s)(\epsilon_1+\epsilon_2-\phi_{s'}+\phi_s)}
\end{aligned}\end{equation}
The above consideration suggests the following form of 
\begin{equation}
\begin{aligned}
Z^{\text{\sf{sym}}}_{\text{\sf{vec}}}(\vec{a},\vec{Y})\equiv \det{}' v=\prod_{\alpha,\beta=1}^{2}
    \prod_{s\in {}^{\diamondsuit}Y_{\alpha}(\beta)}E\bigl(a_{\alpha}-a_{\beta},Y_{\alpha},Y_{\beta}\bigl|s\bigr)
(Q-E\bigl(a_{\alpha}-a_{\beta},Y_{\alpha},Y_{\beta}\bigl|s\bigr)),
\label{determinant}
\end{aligned}
\end{equation}
where the region ${}^{\diamondsuit}Y_{\alpha}(\beta)$  is defined (see \eqref{YT1}) as 
\begin{equation}
{}^{\diamondsuit}Y_{\alpha}(\beta)=
\bigl\{(i,j) \in Y_{\alpha} \bigl| P\bigl(k'_j(Y_\alpha)\bigl)\neq P\bigl(k_i(Y_\beta)\bigl)\bigl\},
\end{equation}
or, in other words, the cells having different parity of the leg- and arm-factors. 

We conjecture the following relation between $\mathbb{Z}_2$ instanton partition function for the pure
gauge situation evaluated on some given component $\mathcal{M}_{\text{sym}}(N_{+},N_{-})$   and super Liouville
conformal blocks in the Whittaker limit~\eqref{whittsuperblock}: 
 \begin{equation}
\begin{aligned}
\label{N1-conjecture}
\sum_{N=0,1,\dots} q^N \sum_{\vec{Y},\substack{N_{+}(\vec{Y})=N\\N_{-}(\vec{Y})=N}} 
\frac{1}{Z^{\text{\sf{sym}}}_{\text{\sf{vec}}}(\vec{a},\vec{Y})}
\;\;\;\;\;\;\; &{=}
\;\;\; F_0(\Delta(a)|q)\;,\\
\sum_{N=\frac 12,\frac 32,\dots} q^N \sum_{\vec{Y},\substack{N_{+}(\vec{Y})=N+\frac 12\\N_{-}(\vec{Y})=N-\frac 12}}\frac{1}{Z^{\text{\sf{sym}}}_{\text{\sf{vec}}}(\vec{a},\vec{Y})}
\;\;\; &{=}
\;\;\; F_1(\Delta(a)|q)\;.
\end{aligned}
\end{equation}

\section{Lowest levels calculations and analytic properties}\label{checks}

We performed explicit calculations of the instanton partition function up to $5$-instantons contribution. 
The results for the norm of the Whittaker vector, which follows from our conjecture \eqref{N1-conjecture}, 
agree with the results \eqref{lowestcoef} derived in section 3. Below we illustrate 
the results for levels $1/2,1,3/2$.
\begin{itemize}
\item {\bf One-instantons contribution} \end{itemize}
In this simple case there are only two pairs of Young diagrams 
\begin{equation}
(Y_1,Y_2)=(\{1\},\{\varnothing \}) \,\,\,\,\text{and}\,\,\,\,(Y_1,Y_2)=(\{\varnothing \},\{1\}) .
\end{equation}
Moreover there is no need to consider the second pair separately since interchanging $Y_1$ and $Y_2$ leads to the 
same determinant with $a$ replaced by $-a$. Taking into account \eqref{determinant} one easily finds
\begin{equation}
\det{}' v(\{1\},\{\varnothing \})=-2a(2a+\epsilon_1+\epsilon_2).
\end{equation}
Thus for one-instantons contribution
\begin{eqnarray}
\sum_{\vec{Y},\substack{N_{+}(\vec{Y})=1\\N_{-}(\vec{Y})=0}}\frac{1}{Z^{\text{\sf{sym}}}_{\text{\sf{vec}}}(\vec{a},\vec{Y})}=\frac{1}{-2a(2a+\epsilon_1+\epsilon_2)}+
\frac{1}{2a(-2a+\epsilon_1+\epsilon_2)}.
\end{eqnarray}
Eq.~\eqref{N1-conjecture} offers the following answer for the coefficient
\begin{equation}
F^{(1/2)}(\Delta)=\langle \frac 12|\frac 12 \rangle=\frac{4b^2}{(1-2a b+b^2)(1+2 a b+b^2)},
\end{equation}
which coincides with \eqref{lowestcoef}.
\begin{itemize}
\item {\bf Two-instantons contribution} \end{itemize}
There are five fixed points in this case.
\begin{equation}
\begin{aligned}
(Y_1,Y_2)=(\{2\},\{\varnothing \}),\,&\,(Y_1,Y_2)=(\{\varnothing\},\{2\}),\\
(Y_1,Y_2)=(\{1,1\},\{\varnothing \}),\,&\,(Y_1,Y_2)=(\{\varnothing \},\{1,1\}),\\ 
(Y_1,Y_2)=&(\{1\},\{1\}).
\end{aligned}
\end{equation}
Now it is sufficient to consider only the first and the last pairs. The remaining pairs can be   
obtained from the first one by means of the interchanging and the transposition of the Young tableaux
(the second operation corresponds to the interchanging $\epsilon_1 \leftrightarrow \epsilon_2$).
The determinants are
\begin{equation}
\begin{aligned}
&\det{}' v(\{2\},\{\varnothing \})=4a\epsilon_1(\epsilon_1-\epsilon_2)(2a+\epsilon_1+\epsilon_2),\\
&\det{}' v(\{1\},\{1 \})=1.
\end{aligned}
\end{equation}
The two-instantons contribution is
\begin{equation}
\begin{aligned}
\sum_{\vec{Y},\substack{N_{+}(\vec{Y})=1\\N_{-}(\vec{Y})=1}}&\frac{1}{Z^{\text{\sf{sym}}}_{\text{\sf{vec}}}(\vec{a},\vec{Y})}=
\frac{1}{4a\epsilon_1(\epsilon_1-\epsilon_2)(2a+\epsilon_1+\epsilon_2)}+
\frac{1}{4a\epsilon_1(\epsilon_1-\epsilon_2)(2a-\epsilon_1-\epsilon_2)}\\
+&\frac{1}{4a\epsilon_2(\epsilon_2-\epsilon_1)(2a+\epsilon_2+\epsilon_1)}+
\frac{1}{4a\epsilon_2(\epsilon_2-\epsilon_1)(2a-\epsilon_2-\epsilon_1)}.
\end{aligned}
\end{equation}
From Eq.~\eqref{N1-conjecture} one finds
\begin{equation}
F^{(1)}(\Delta)=\langle 1\,|\, 1 \rangle=\frac{b^2}{(1-2a b+b^2)(1+2 a b+b^2)}.
\end{equation}
\begin{itemize}
\item {\bf Three-instantons contribution} \end{itemize}
Fixed points:
\begin{equation}
\begin{aligned}
&(Y_1,Y_2)=(\{3\},\{\varnothing \}),\,\,(Y_1,Y_2)=(\{\varnothing\},\{3\}),\\
&(Y_1,Y_2)=(\{1,1,1\},\{\varnothing \}),\,\,(Y_1,Y_2)=(\{\varnothing \},\{1,1,1\}),\\
&(Y_1,Y_2)=(\{2\},\{1 \}),\,\,(Y_1,Y_2)=(\{1\},\{2\}),\\
&(Y_1,Y_2)=(\{1,1\},\{1 \}),\,\,(Y_1,Y_2)=(\{1 \},\{1,1\}),\\ 
&(Y_1,Y_2)=(\{2,1\},\{\varnothing \}),\,\,(Y_1,Y_2)=(\{\varnothing \},\{2,1\}).
\end{aligned}
\end{equation}
In this case there are three independent determinants 
\begin{equation}
\begin{aligned}
&\det{}' v(\{3\},\{\varnothing \})=4a(2a+2\epsilon_1)\epsilon_1(\epsilon_2-\epsilon_1)
(2a+\epsilon_1+\epsilon_2)(2a+3\epsilon_1+\epsilon_2),\\
&\det{}' v(\{2,1\},\{\varnothing \})=-2a(2a+\epsilon_1+\epsilon_2),\\
&\det{}' v(\{2\},\{1\})=4a(2a+2\epsilon_1)\epsilon_1(\epsilon_2-\epsilon_1)
(2a+\epsilon_1-\epsilon_2)(2a+\epsilon_1+\epsilon_2).
\end{aligned}
\end{equation}
From Eq.~\eqref{N1-conjecture} we derive the following answer
\begin{equation}
\begin{aligned}
&F^{(3/2)}(\Delta)=\langle \frac 32|\frac 32 \rangle=\frac{1}{(-1+2ab-b^2)(1+2ab+b^2)}\times\\
&\times \frac{4b^4(-9-22b^2+4a^2b^2-9b^4)}
{(-1+2ab-3b^2)(-3+2ab-b^2)(3+2ab+b^2)(1+2ab+3b^2)}.
\end{aligned}
\end{equation}

\paragraph{Analysis of the physical poles.}
Another confirmation of our main statement \eqref{N1-conjecture} comes from the analysis of the
conformal block singularities. The conformal block coefficients have poles if $a=\pm \lambda_{m,n}$ 
\cite{HEMN1,Belavin:2007zz}. The residues are given by 
\begin{equation}
 \text{Res} \, F^{(N)}_{a=\pm
\lambda_{m,n}}=(r_{m,n})^{-1}, 
\label{Res_nm_boson}
\end{equation}
where the integers $m$ and $n$, either both even or both odd, should satisfy  $m n= 2N$ and
the coefficients
\begin{equation}
r_{m,n}=2^{1-mn}\prod_{(k,l)\in[m,n]}(kb^{-1}+lb).
\label{rmn}
\end{equation}
Here
\begin{equation}
[m,n]=\{1-m:2:m-1,1-n:2:n-1\}\cup\{2-m:2:m,2-n:2:n\}\setminus(0,0).
\end{equation}

In the  Young diagram decomposition of the conformal block coefficient of the Nth level the poles $a=\pm\lambda_{m,n}$ 
(such that $m n=2N$) appear only in the contributions related to the pairs $(Y_1,Y_2)$, where one of the 
diagram is rectangle with the weight $n$ and the length $m$, and another one is empty. Let us consider
$Y_2=\varnothing$. If $a=-\lambda_{m,n}$ the pole appears in $E(2a,Y_1,\varnothing)$ in the cell $(m,1)$.
One can rewrite $r_{m,n}$ in the form, which is more adopted for the interpretation
in terms of the Young diagrams
\begin{equation}
\label{rmnodd}
\begin{aligned}
r_{m,n}=&\prod_{\substack{s\in Y_1, s\neq(m,1)\\ s-\text{white}}}
[(k-m)b^{-1}+(1-l)b]\prod_{\substack{s\in Y_1\\s-\text{white}}}[Q-((k-m)b^{-1}+(1-l)b)]\\
&\times\prod_{\substack{s\in Y_1\\s-\text{black}}}
[(k-m)b^{-1}+(n+1-l)b]\prod_{\substack{s\in Y_1\\s-\text{black}}}[Q-((k-m)b^{-1}+(n+1-l)b)],
\end{aligned}
\end{equation}
for $(m,n)$ both odd and
\begin{equation}
\label{rmneven}
\begin{aligned}
r_{m,n}=&\prod_{\substack{s\in Y_1, s\neq(m,1)\\ s-\text{black}}}
[(k-m)b^{-1}+(1-l)b]\prod_{\substack{s\in Y_1\\s-\text{black}}}[Q-((k-m)b^{-1}+(1-l)b)]\\
&\times\prod_{\substack{s\in Y_1\\s-\text{black}}}
[(k-m)b^{-1}+(n+1-l)b]\prod_{\substack{s\in Y_1\\s-\text{black}}}[Q-((k-m)b^{-1}+(n+1-l)b)],
\end{aligned}
\end{equation}
for $(m,n)$ both even.
We note that 
\begin{equation}
\begin{aligned}
&[(k-m)b^{-1}+(1-l)b]=E(-2\lambda_{m,n},Y_1,\varnothing),\\
&[(k-m)b^{-1}+(n+1-l)b]=E(-2\lambda_{m,n},Y_1,Y_1),
\end{aligned}
\end{equation}
so that this form of the residue $r_{m,n}$ almost coincide with the expression which follows from \eqref{determinant}.
It remains to verify that the regions in
\eqref{rmnodd} and \eqref{rmneven} coincide with the region ${}^{\diamondsuit}Y_{\alpha}(\beta)$, i.e. form a subset
of cells with different parity of the leg- and arm-factors. 
Consider first $(Y_1,\varnothing)$, which corresponds to the first lines
in \eqref{rmnodd} and \eqref{rmneven}. The leg-factor $L_{Y_1}(s)=m-i$ and the arm-factor $A_{\varnothing}(s)=-j$.
If $s$ is white, the coordinates $i$ and $j$ have the same parity. Hence, for odd $m$,  
$L_{Y_1}(s)$ and $A_{\varnothing}(s)$ are of different parity, as it should be. Similarly, if $s$ is black
and $m$ is even, $L_{Y_1}(s)$ and $A_{\varnothing}(s)$ also have different parity. Finally, consider
the second lines in \eqref{rmnodd} and \eqref{rmneven}, related to the pair $(Y_1,Y_1)$, 
then $L_{Y_1}(s)=m-i$ and $A_{Y_1}(s)=n-j$. Similar arguments shows that only black cells satisfy necessary 
requirement, that is belong to ${}^{\diamondsuit}Y_{\alpha}(\beta)$.

\section{Conclusion}
In this paper we formulate and perform some tests of the following statement. 
The subspace of the $SU(2)$ moduli space which consists of $\mathbb{Z}_2$ symmetric instanton solutions
is related to the $\cN=1$ super Liouville theory. Namely, the conformal block function in the Whittaker
limit coincides with the instanton partition function evaluated by means of the 
localization technique in the reduced moduli space . Our proposal generalizes the AGT relation between the 
ordinary Liouville theory and $SU(2)$ quivers. The idea comes from the observation that the algebra acting 
on the cohomologies of $\mathbb{Z}_2$ symmetric instanton varieties in the conformal limit is 
$\assalgebra{A}=\hat{gl}_2(2)\times \assalgebra{NSR}$ instead of $\assalgebra{H}\times \assalgebra{V}ir$ in the case 
of the ordinary AGT correspondence. 

Further study of the proposed relation is clearly necessary. In particular,
it would be nice to find the orthogonal basis, which consists of the eigenvectors of some commuting
subalgebra of $\assalgebra{A}$, as it was done in \cite{ALTF,BB}. 
Is is interesting also to derive the representation for the four-point super conformal block. We are going to do this
in the next publication. Another open question is what kind of gauge theory is behind the
modified instanton moduli space discussed in this paper. Finally, it is
clearly intriguing to generalize our proposed construction to the action
of other possible finite groups acting on the instanton moduli, in
particular, to the $\mathbb{Z}_m$ group action.

\section*{Acknowledgments}
The authors are grateful to Kostya Alkalaev, Alexander Belavin, Misha Bershtein and Alexei Litvinov for many 
fruitful discussions. The research was held within the framework of the Federal 
programs ``Scientific and Scientific-Pedagogical 
Personnel of Innovational Russia'' on 2009-2013 (state contracts No. P1339) and supported by the joint RFBR-CNRS
grant No. 09-02-93106 and RFBR grant No.11-01-12023-ofi-m-2011.

\providecommand{\href}[2]{#2}\begingroup\raggedright\endgroup


\begin{thebibliography}{10%
0}
\addtolength{\baselineskip}{-3pt}
\addtolength{\parskip}{-1pt}


\bibitem{Alday:2009aq}
L.~F. Alday, D.~Gaiotto, and Y.~Tachikawa, {\it {Liouville Correlation
  Functions from Four-dimensional Gauge Theories}},  {\em Lett. Math. Phys.}
  {\bf 91} (2010) 167--197, [\href{http://xxx.lanl.gov/abs/0906.3219}{{\tt
  arXiv:0906.3219}}].


\bibitem{Belavin:1984vu}
A.~A. Belavin, A.~M. Polyakov and A.~B. Zamolodchikov, {\it Infinite conformal
  symmetry in two-dimensional quantum field theory},  {\em Nucl. Phys.} {\bf
  B241} (1984) 333--380.

\bibitem{Alday:2010vg}
L.~F. Alday and Y.~Tachikawa, {\it {Affine SL(2) conformal blocks from 4d gauge
  theories}},  {\em Lett. Math. Phys.} {\bf 94} (2010) 87--114,
  [\href{http://xxx.lanl.gov/abs/1005.4469}{{\tt arXiv:1005.4469}}].


\bibitem{Wyllard:2009hg}
N.~Wyllard, {\it {$A_{N-1}$ conformal Toda field theory correlation functions
  from conformal $N=2$ $SU(N)$ quiver gauge theories}},  {\em JHEP} {\bf 11}
  (2009) 002, [\href{http://xxx.lanl.gov/abs/0907.2189}{{\tt
  arXiv:0907.2189}}].

\bibitem{MM} 
A.~Mironov, A.~Morozov, {\it{On AGT relation in the case of U(3)}},
 {\em Nucl. Phys.}   {\bf B825} (2010) 1--37,
[\href{http://xxx.lanl.gov/abs/0908.2569}{{\tt arXiv:0908.2569}}].



\bibitem{Taki} 
M.~Taki, {\it{On AGT Conjecture for Pure Super Yang-Mills and W-algebra}},
[\href{http://xxx.lanl.gov/abs/0912.4789}{{\tt arXiv:0912.4789}}].




\bibitem{Lusztig} G. Lusztig, {\it{Quivers, perverse sheaves, and quantized enveloping algebras}},
{\em J. Amer. Math. Soc.} {\bf 4} (1991), no. 2, 365–-421.

\bibitem{Nakajima1} H.~Nakajima, {\it{Instantons on ALE spaces, quiver varietie, and Kac-Moody algebras}},
{\em Duke Math.} {\bf 76} (1994) 365--416. 


\bibitem{Nakajima} H.~Nakajima and K.Yoshioka, {\it {Instanton counting on blowup. I. 4-dimensional pure gauge 
theory}}, {\em Invent. Math} {\bf 162} (2005) no.2, 313--355,
[\href{http://xxx.lanl.gov/abs/math/0306198}{{\tt math/0306238}}].

\bibitem{Feigin1} B. Feigin, A. Tsymbaliuk, {\it{Heisenberg action in the equivariant K-theory of Hilbert
schemes via scuffle algebra}},  
[\href{http://xxx.lanl.gov/abs/0904.1679}{{\tt arXiv:0904.1679}}].

\bibitem{Nagao} K. Nagao, {\it{K-theory of quiver varieties, q-Fock space and nonsymmetric Macdonald polynomials}},
[\href{http://xxx.lanl.gov/abs/0709.1767}{{\tt arXiv:0709.1767}}].

\bibitem{Flume:2002az}
R.~Flume and R.~Poghossian, {\it {An algorithm for the microscopic evaluation
  of the coefficients of the Seiberg-Witten prepotential}},  {\em Int. J. Mod.
  Phys.} {\bf A18} (2003) 2541,
  [\href{http://xxx.lanl.gov/abs/hep-th/0208176}{{\tt hep-th/0208176}}].

\bibitem{B} J.-M. Bismut, Localization Formulas, {\it{Superconnections
and the Index Theorem of Families}}, {\em Commun.Math.Phys.} {\bf 103} (1986), 127-166.


\bibitem{BGV}N. Berline, E. Getzler and M. Vergne,
{\it{Heat Kernels and Dirac Operators}}, Springer, Berlin, (1996).

\bibitem{Nekrasov:2002qd}
N.~A. Nekrasov, {\it {Seiberg-Witten Prepotential From Instanton Counting}},
  {\em Adv. Theor. Math. Phys.} {\bf 7} (2004) 831--864,
  [\href{http://xxx.lanl.gov/abs/hep-th/0206161}{{\tt hep-th/0206161}}].


\bibitem{WhittakGaiot} D. Gaiotto, {\it{Asymptotically free N=2 theories and irregular conformal blocks}},
[\href{http://xxx.lanl.gov/abs/0908.0307 }{{\tt arXiv:0908.0307}}].

\bibitem{MMM} A.~Marshakov, A.~Mironov, A.~Morozov, 
{\it{On non-conformal limit of the AGT relations}}, 
{\em Phys.Lett.} \textbf{B682} (2009) 125--129,
[\href{http://xxx.lanl.gov/abs/0909.2052 }{{\tt arXiv:0909.2052}}].


\bibitem{Kostant} B. Kostant, {\it{On Whittaker vectors and representation theory}}, 
{\em Invent. Math.} {\bf 48} (1978) 101--184.


\bibitem{Polyakov}A. Polyakov. {\it{Quantum geometry of fermionic strings}},
{\em Phys.Lett.} \textbf{B103} (1981) 211--213.

\bibitem{Belavin:2007gz}
  A.~Belavin, V.~Belavin, A.~Neveu and A.~Zamolodchikov,
  {\it {Bootstrap in Supersymmetric Liouville Field Theory. I. NS Sector}},
 {\em Nucl. Phys.}  {\bf B784} (2007) 202,
[\href{http://xxx.lanl.gov/abs/hep-th/0703084}{{\tt hep-th/0703084}}].

\bibitem{Hadasz:2006sb}
    L.~Hadasz, Z.~Jask\'olski and P.~Suchanek,
    {\it{ Recursion representation of the Neveu-Schwarz superconformal block}},
    {\em JHEP} {\bf 03} (2007) 032,
[\href{http://xxx.lanl.gov/abs/hep-th/0611266}{{\tt hep-th/0611266}}].

\bibitem{Belavin:2007zz}
  V.~A.~Belavin,
  {\it {N=1 SUSY conformal block recursive relations}},
  {\em Theor. Math. Phys.}  {\bf 152} (2007) 1275,
[\href{http://xxx.lanl.gov/abs/hep-th/0611295}{{\tt hep-th/0611295}}].


\bibitem{Belavin:2007eq}
  V.~A.~Belavin,
  {\it {On the N = 1 super Liouville four-point functions}},
  {\em Nucl. Phys.}   {\bf B798} (2008) 423,
[\href{http://xxx.lanl.gov/abs/0705.1983}{{\tt ArXiv:0705.1983}}].


\bibitem{Hadasz:2007nt}
  L.~Hadasz, Z.~Jaskolski and P.~Suchanek,
{\it {Elliptic recurrence representation of the N=1 Neveu-Schwarz blocks}},
 {\em Nucl. Phys.}   {\bf B798} (2008) 363,
[\href{http://xxx.lanl.gov/abs/0711.1619}{{\tt arXiv:0711.1619}}].

\bibitem{adhm} M. Atiyah, V. Drinfeld, N. Hitchin, Yu. Manin, {\it {Constractions of instantons}},
{\em Phys. Lett.} {\bf A65} (1978) 185.


\bibitem{BPSz} A. Belavin, A. Polyakov, A. Schwartz, Yu. Tyupkin, 
{\it{Pseuvdoparticle solutions of the Yang-Mills equations}},
{\em Phys. Lett.} {\bf B59} (1975) 85--86. 


\bibitem{BZakh} A. Belavin, V. Zakharov, {\it{Yang-Mills equations as inverse scattering problem}},
{\em Phys. Lett.} {\bf B73} (1978) 53--57.


\bibitem{aw} M. Atiyah, R. Ward, {\it{Instantons and algebraic geometry}},
{\em Comm. Math. Phys.} {\bf 55} (1977) 117--127.

\bibitem{dm} V. Drinfeld, Yu. Manin, {\it{ Descriptions of instantons}},
{\em Comm. Math. Phys.} {\bf 6} (1978) 177--192.


\bibitem{DKM'} N. Dorey, T. Hollowood, V. Khoze, M. Mattis, {\it {The Calculus of Many Instantons}},
[\href{http://xxx.lanl.gov/abs/hep-th/0206063}{{\tt hep-th/0206063}}].

\bibitem{HEMN1} A. Belavin, Al. Zamolodchikov, {\it {Higher Equations of Motion in $\cN=1$ SUSY
Liouville field theory}}, 
{\em JETP lett.} {\bf 84} (2006) 496--502
[\href{http://xxx.lanl.gov/abs/hep-th/0610316}{{\tt hep-th/0610316]}}]

\bibitem{ALTF}
V.~Alba, V.~Fateev, A.~Litvinov and G.~Tarnopolsky,
{\it {On combinatorial expansion of the conformal blocks arising from
AGT conjecture}}, [\href{http://xxx.lanl.gov/abs/1012.1312}{{\tt arXiv:1012.1312}}].


\bibitem{BB}
A.~Belavin, V.~Belavin, {\it {AGT conjecture and Integrable structure
of Conformal field theory for $c=1$}}, 
{\em Nucl. Phys.} {\bf B850} (2011) 199--213,
[\href{http://xxx.lanl.gov/abs/1102.0343}{{\tt arXiv:1102.0343}}].


\end{thebibliography}
\end{document}